\documentclass[a4paper, 10pt]{article}

\pdfoutput = 0

\usepackage[utf8]{inputenc}
\usepackage[T1]{fontenc}
\usepackage{fixltx2e}
\usepackage{graphicx}
\usepackage{longtable}
\usepackage{float}
\usepackage{wrapfig}
\usepackage{rotating}
\usepackage[normalem]{ulem}
\usepackage{amsmath}
\usepackage{textcomp}
\usepackage{marvosym}
\usepackage{wasysym}
\usepackage{amssymb}
\usepackage{RR}
\usepackage[numbers]{natbib}
\usepackage{microtype}
\usepackage[affil-it]{authblk}
\author{Ludovic Courtès\thanks{\texttt{ludovic.courtes@inria.fr}}}
\affil{Inria (Bordeaux, France)}
\RRetitle{Design and Implementation of rowe, a Web-Friendly Communication Library}
\RRtitle{Conception et mise en œuvre de rowe, une bibliothèque de communication orientée Web}
\RRauthor{Ludovic Courtès}
\RCBordeaux
\RRprojet{Indes}
\RRkeyword{communication library, websocket, JSON}
\RRmotcle{bibliothèque de communication, websocket, JSON}
\RTNo{452}
\date{\today}
\title{Design and Implementation of rowe, a Web-Friendly Communication Library}
\begin{document}

\RRabstract{
The INDES project-team of Inria has been developing HOP, a multi-tier
language for Web programming.  As part of the RAPP FP7 European project,
the team has set out to use HOP as the \emph{lingua franca} of the robotics
applications developed within that project.  Part of the challenge lies
in the integration of existing robotics code, written using ROS or
custom libraries, with HOP-based application.

This document reports on the implementation of rowe, a communication
library designed the fill the gap between low-level robotics C
components on one hand, and other C, C++, ROS, or HOP components on the
other.  The library aims to be a lightweight, high-performance,
``Web-friendly'' communication library.  It implements a socket-like
interface that allows programs to exchange JSON objects over WebSockets.
We describe the rationale, design, and implementation of rowe.
}

\RRresume{L'équipe-projet Inria INDES développe HOP, un langage à
  plusieurs niveaux pour la programmation Web.  Dans le cadre du projet
  européen FP7 RAPP, l'équipe s'est donné pour objectif d'utiliser HOP
  comme \textit{lingua franca} des applications robotiques développées
  dans le projet.  Un des défis à relever est l'intégration de code
  robotique existant, utilisant ROS ou des bibliothèques dédiées, avec
  des applications en HOP.

  Ce document décrit rowe, une bibliothèque de communication visant à
  combler le vide entre d'un côté des composants robotiques bas niveau
  écrits en C, et d'un autre côté des composants écrits en C++, ROS ou
  HOP.  L'objectif est de fournir une bibliothèque de communication
  légère, haute performance et qui s'intègre facilement à
  l'environnement Web.  Elle met en œuvre une interface de type
  \textit{socket} permettant d'échanger des objets JSON sur des
  WebSockets.  Nous décrivons les motivations, les choix de conception,
  et la mise en œuvre de rowe.
}

\makeRT
\clearpage
\tableofcontents

\section{Introduction}
\label{sec-1}

RAPP\footnote{See the RAPP Web site at \url{http://rapp-project.eu/}.} is a project of the European Commission's Seventh Framework
Program (FP7).  It seeks to deliver a software platform including
robotic applications for social inclusion.  The goal is to deliver
robotic applications (or ``RApps'') for a variety of a robots---ranging
from humanoid robots, to an instrumented walking aid for elderly
people---that will often need to interact with Web services, for
instance to access user profile information.  Those applications
typically use a variety of programming languages and technologies, such
as C, C++, Python, ROS\footnote{ROS Web site at \url{http://www.ros.org/}.}, and JavaScript.

HOP is a Web programming framework \cite{serrano06:hop} that makes it
easy to use the protocols and formats of the Web: HTTP,
WebSocket \cite{websocketrfc11}, JSON, etc.  HOP also comes with a
package management tool, called Hz, that simplifies the installation of
HOP applications.  HOP was chosen as the tool to build the
infrastructure for RAPP's application store and on-line services, as
well as the tool to connect application components
together \cite{rapp14:architecture}.

To that end, we developed two pieces of software: a HOP interface to ROS, and
rowe, a C library to connect with embedded robotics software that does not use
ROS---such as the Assistive Navigation Guide (ANG) family of walking aids
developed by Inria's Héphaïstos team\footnote{See an overview of the ANG family of walking aids at
\url{https://pal.inria.fr/research/themes/rehabilitation-transfer-and-assistance-in-walking/walking-aids/}.}.

The next section describes the goals and design choices that we set out
for rowe.  Section \ref{sec-3} gives and overview of its
implementation.  Section \ref{sec-4} concludes.

\section{Rationale}
\label{sec-2}

The rowe library aims to connect C robotics software with components
that communicate using the ROS protocols or Web protocols.

\subsection{Design Goals}
\label{sec-2-1}

We set out a number of design goals for rowe:

\begin{enumerate}
\item Robotics software using rowe is going to send status updates to
other RAPP components at a possibly high rate (for instance, the
speed and location of the robot, similar to ROS ``topics''), and it
must receive and process requests from other components in a timely
fashion (such requests may include an emergency stop, for instance,
similar to ROS services.)  Thus, rowe must guarantee low latency
and high throughput.

\item Programs using rowe are meant to be connected primarily with HOP
programs, so rowe must be a ``native speaker'' of the Web protocols
and formats.

\item It must be possible using rowe to exchange typed and structured
messages, such as strings, numbers, records, lists, and so on.

\item The application programming interface (API) of rowe should match
the programming style and expectations of low-level robotics
developers.  Informally, that means that it should be as little
disruptive as possible.
\end{enumerate}

Being a native of the Web meant that rowe's transport layer should be
based on HTTP, which also has the advantage of being usually allowed
traffic through firewalls.  Of course using HTTP alone to exchange
messages, for instance in a ReST fashion, would incur too much overhead:
HTTP connections would regularly need to be instantiated, which
unacceptably increases latency, and HTTP \texttt{GET} requests may incur too
much bandwidth overhead.

For that reasons, we chose to use WebSockets as the transport
later \cite{websocketrfc11}.  WebSockets is an HTTP extension that
provides a reliable, bidirectional communication channel that can be
used to transfer arbitrary payloads, similar to TCP.

JSON, for JavaScript Object Notation \cite{ecma13:json}, came up as the
obvious choice for the message format.  It meets our requirements as a
mechanism to encode structured and typed messages, it is the natural way to
represent data in HOP programs, and has efficient parsers and serializers.

The last design goal is more subjective.  In our view, matching the
programming style of low-level robotics developers meant a few things.
First, the API should be usable in single-threaded programs.  Second, it
should not expose a full-blown event loop framework as commonly found in
object-oriented libraries such as GLib.  Those frameworks are generally
complex, and they impose \emph{inversion of control} (IoC) through a heavy
use of callbacks: that essentially forces developers to write in
continuation-passing style (CPS), which is both verbose and difficult to
work with.  Instead, we want to allow a direct programming style.  This
has been the main choice driving the design of the programming
interface.

\subsection{Programming Interface}
\label{sec-2-2}

The bulk of rowe's programming interface has purposefully been kept
minimal and simple.

In rowe version 1, connections are modeled by an \emph{endpoint}.  A rowe
program can only be connected to one peer at a time (see Section
\texttt{Conclusion} for a discussion and desired changes to this approach.)  A
rowe program can be an HTTP server:

\begin{verbatim}
struct rowe_enpoint *endpoint;
endpoint = rowe_open_local_endpoint (8080);
\end{verbatim}

or it can be an HTTP client:

\begin{verbatim}
endpoint = rowe_open_remote_endpoint ("hop.inria.fr", 8080);
\end{verbatim}

The API to send and receive messages is the same regardless of whether
the program is a server or a client.

Messages are JSON objects, as implemented by the JSON-C library\footnote{The JSON-C library: \url{https://github.com/json-c/json-c/wiki}.}
sent to an endpoint using the \texttt{rowe\_send} function.  Callers can specify
a time-to-live (TTL) for the message: if no peer was connected after the
TTL has expired, the message is discarded and not sent.  This is useful
for periodic messages such as updates on the robot's status, akin to ROS
topics: an old update is not valuable and can be discarded, allowing the
peer to get fresher updates instead.

\begin{verbatim}
extern int rowe_send (const struct rowe_endpoint *endpoint,
		      const struct json_object *obj,
		      long ttl);
\end{verbatim}

The \texttt{rowe\_send} function is synchronous and blocks until the message has
either been sent, or has been discarded.  Alternately, the
\texttt{rowe\_async\_send} function is non-blocking, any may typically be used
when sending messages containing status updates.

Similarly, the \texttt{rowe\_receive} functions blocks until a message is
received or the user-specified timeout has expired, and returns a
pointer to a \texttt{json\_object} structure or \texttt{NULL}.

Programs may also perform remote procedure calls (RPCs), using the
\texttt{rowe\_invoke} function:

\begin{verbatim}
extern struct json_object *
rowe_invoke (const struct rowe_endpoint *endpoint,
	     struct json_object *obj, long timeout);
\end{verbatim}

The function sends the given JSON object, which denotes a procedure
invocation, blocks until a reply has been received, and returns it,
unless the given timeout has expired.  The actual format of the JSON
object representing the procedure call is at the user's discretion.  An
example JSON-formatted service invocation may look like this:

\begin{verbatim}
{
  "service": "add-two-numbers",
  "a": 38,
  "b": 4
}
\end{verbatim}

Lastly, rowe programs can reply to RPCs, using the \texttt{rowe\_reply} function
or using \texttt{rowe\_async\_reply}, its non-blocking counterpart.

To facilitate the creation of JSON objects representing key/value
associations, the \texttt{rowe\_message} convenience function is provided.  For
instance, the message shown above may be instantiated with the following
call:

\begin{verbatim}
struct json_object *invocation;

invocation = rowe_message ("service",
			   json_type_string, "add-two-numbers",
			   "a", json_type_int, 38,
			   "b", json_type_int, 4, NULL);
\end{verbatim}

\section{Implementation}
\label{sec-3}

The implementation of the above API goes along the following lines.

\subsection{Service Thread}
\label{sec-3-1}

rowe builds upon the JSON-C\footnotemark[4]{} and libwebsockets\footnote{The libwebsockets library: \url{http://libwebsockets.org/}.} libraries.
Since it does not expose an event loop interface, the actual event loop
runs in a dedicated service thread, which is spawned when the endpoint
is opened.  The service thread polls for connection requests and for
``in'' and ``out'' events on open connections.

The service thread adds incoming messages on a queue that is checked by
functions such as \texttt{rowe\_receive}.  When \texttt{rowe\_send} and similar
functions are called from the user thread, they add the given message to
an outgoing message queue, which the service thread checks when the
connection is ready to accept outgoing messages.  When the service
thread accesses the outgoing message queue, it deletes any messages
whose TTL has expired.

The \texttt{rowe\_async\_send} function is the simplest: it just adds a message
to the outgoing message queue and returns immediately.  Conversely, the
\texttt{rowe\_send} and \texttt{rowe\_receive} functions need to synchronize with the
service thread.  \texttt{rowe\_receive} checks for message in the incoming
message queue; when the message queue is empty, it waits on a condition
variable associated with it.  \texttt{rowe\_send} works by passing a
\emph{notification} object, which essentially bundles together a condition
variable and a return value, which the service thread notifies when the
message is discarded due to TTL expiration, or once it has been sent.

\subsection{Remote Procedure Calls}
\label{sec-3-2}

RPC replies need special treatment: when \texttt{rowe\_invoke} is used,
unrelated messages may be received after the invocation message has been
sent and before the reply has been received; yet, \texttt{rowe\_invoke} must
return the RPC reply, not another message that happened to be received
first.

To address that, rowe takes several steps.  First, it requires
invocation messages to be JSON objects (key/value associations) and,
upon invocation, it automatically adds them a \texttt{message\_id} entry whose
value is a unique identifier, allowing the invocation to be
distinguished from other invocations of the same remote procedure.  RPC
replies must also be JSON objects, and they must have a \texttt{in\_reply\_to}
entry whose value is the \texttt{message\_id} of a previous invocation message.

Second, rowe maintains a table that allows it to match RPCs with
replies, and to wake up the user thread that is waiting in
\texttt{rowe\_invoke}.  The table is essentially a list of pairs of \texttt{message\_id}
values and corresponding notification object that allows the user
thread, which may be waiting in \texttt{rowe\_invoke}, to be woken up.

\subsection{Time-to-Live}
\label{sec-3-3}

Since the service thread must periodically removed expired messages both
from the incoming and the outgoing message queues, this operation must
be efficient.  To that end, the message queue structure is (1) a
doubly-linked list, which provides for constant-time deletion, and (2)
it can be viewed both as a queue (FIFO) and as a list of messages
ordered by expiration date, which makes it more efficient.

\section{Summary}
\label{sec-4}

The rowe library provides a simple programming interface for connected
components.  It is well suited for the RAPP project where it allows
low-level robotics software to communicate with HOP programs using Web
protocols, and with good performance, notably on low-end embedded
ARM-based devices.

As of this writing, version 2 of rowe is being developed.  The main goal
is to allow users to distinguish between an endpoint and an established
connection, and to support connections with multiple peers.

The rowe library is free software, available from
\url{https://github.com/rapp-project/rowe} and from
\url{ftp://ftp-sop.inria.fr/indes/rapp/rowe/}.

\bibliographystyle{plain}
\bibliography{rowe}
\end{document}